\documentclass[epj]{svjour}
\usepackage{graphics}
\usepackage{epstopdf}
\usepackage{amsbsy}
\usepackage{amsmath}
\usepackage{amssymb}

\begin{document}
\title{Indirect interactions of membrane-adsorbed cylinders}
\author{Thomas R.\ Weikl}
\institute{Max-Planck-Institut f\"ur Kolloid- und Grenzfl\"achenforschung,
 14424 Potsdam, Germany}
\date{}
\abstract{Biological and biomimetic membranes often contain
aggregates of embedded or adsorbed macromolecules. In this article, 
the indirect interactions of cylindrical objects adhering  to a planar
membrane are considered theoretically. The adhesion of the 
cylinders causes a local perturbation of the equilibrium membrane 
shape, which leads to membrane-mediated interactions.
For a planar membrane under lateral tension, the
interaction is repulsive for a pair of cylinders adhering to the same side
of the membrane, and attractive for cylinders adhering at 
opposite membrane sides. For a membrane in an external
harmonic potential, the interaction of adsorbed cylinders is
always attractive and increases if forces perpendicular to the
membrane act on the cylinders.
 }
\maketitle
%

\section{Introduction}

Biological membranes consist of a multi-component lipid bilayer with a 
variety of embedded or adsorbed macromolecules  such as proteins
\cite{alberts94,lipowsky95}. In recent years, experiments revealed a
complex lateral architecture of these membranes which contain domains 
or `rafts'  of different molecular composition. The domains often serve
important biological functions in signaling \cite{simons97}, budding 
\cite{schekman96},  or cell adhesion \cite{monks98,grakoui99}.

Lateral phase separation and domain formation has also
been observed for biomimetic membranes,  which are composed of only 
a few different molecules. In principal, the phase separation may either be caused (i) by a demixing of the lipid bilayer
\cite{keller98,dietrich01}, or (ii) by the aggregation of embedded or
adsorbed macromolecules \cite{sackmannGroup,koltover99}. To understand
the latter, membrane-mediated interactions between macromolecules have
been studied intensively. They can be divided into static and dynamic 
interactions. {\it Dynamic}, or Casimir-like, interactions arise from the
suppression of membrane shape fluctuations by embedded macromolecules
such as rigid inclusions
\cite{goulian93,netz,park96,golestanian96,dommersnes99,helfrich01,tw1}
or specific receptors or stickers \cite{bruinsma94,tw2}. 
{\it Static} interactions are due to a perturbation of the equilibrium bilayer
structure or equilibrium membrane shape by the embedded or adsorbed
macromolecules. Examples for such macromolecules are trans-membrane
proteins \cite{dan94,fournier98,may99,harroun99,sens00}
and adsorbed molecules \cite{schiller00} causing a perturbation of the equilibrium
membrane thickness,  as well as conical or anisotropic inclusions \cite{goulian93,tw98,dommersnes98,kim98_99,sintes98,dommersnes99,dommersnes02} 
and membrane-anchored polymers \cite{breidenich00,bickel00_01} which cause 
a perturbation of the equilibrium membrane curvature.

In this article, the static interactions of parallel cylindrical objects adhering to a planar membrane are considered. A single such cylinder 
 has been recently studied in Ref.~\cite{boulbitch02}.
 The cylinders are characterized by their radius $R$ and a
favorable adhesion energy per area $U$ which is balanced in equilibrium 
by the elastic energy of the membrane shape deformation. 
Examples for such objects are cylindrical viruses or coated latex cylinders similar
to the beads considered in Ref.~\cite{koltover99}.
The membrane shape around an adsorbed cylinder is similar to the shape
around an elongated wedge-shaped inclusion characterized by its diameter $D$ and
angle $\alpha$. However, the pair interaction energies are
different since the contact area of  parallel adsorbed cylinders depends on the
distance $L$ between the cylinders, while the diameter $D$ and angle $\alpha$ 
of wedge-shaped inclusions are constant.

\section{General model and geometry}

In the absence of adhering objects, the membrane considered here is
planar and constrained into the $x$-$y$ plane either by a lateral
tension $\sigma$ (section 3) or by a harmonic potential (section 4).
The adhering cylinders are characterized by their radius $R$ and
adhesion energy $U$ per area. The cylinders are assumed to be
parallel and to be much longer than their distance $L$. The membrane
profile then is approximated by the one-dimensional function $h(x)$
measuring the deviation out of  the $x$-$y$ plane. Here, $x$ is the cartesian
coordinate perpendicular to the cylinder axes. The membrane bending
energy per area is given by $\frac{1}{2}\kappa h''(x)^2$ where 
$\kappa$ is the bending rigidity  and $h''(x)$ is the total membrane 
curvature.

\begin{figure}[h]
\resizebox{\columnwidth}{!}{\includegraphics{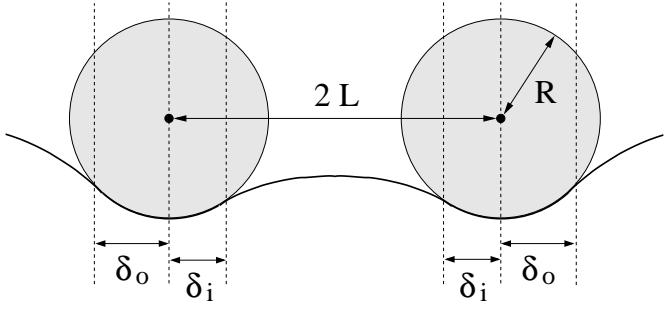}}
\caption{Sketch of a membrane with two parallel cylinders adhering at the
same side. \label{sketch}}
\end{figure}

In order to determine the equilibrium membrane energy, we first determine
the shape and energy of the membrane as a function of the contact area
with the cylinder(s). For a single cylinder with axis located at $x=0$, the
contact area is given by $|x| < \delta_o$. For a pair of cylinders with axes
located at $x=\pm L$, the contact areas are given by 
$L-\delta_\iota < |x| < L+\delta_o$ (see Fig.~1). In equilibrium, the system is
`free' to choose its contact area. Therefore, we finally minimize the free energy
with respect to the contact parameters $\delta_o$, $\delta_\iota$ and the
deviation $h_o$ of the cylinders out of the $x$-$y$ plane. In the
case of a single cylinder, the contact parameter $\delta_o$ divides
the membrane into the contact region and the two `outer' membrane regions
with $|x| > \delta_o$. In the case of two cylinders, we have two contact
regions, an `inner' membrane region with $|x|< L-\delta_\iota$ between the
cylinders, and two `outer' membrane regions with  $|x|> L+\delta_o$.

\section{Membrane under lateral tension}

In the presence of a lateral tension $\sigma$, the 
elastic energy of the membrane can be written as 
\begin{equation}
G =\int \left(\frac{\kappa}{2} h''(x)^2 +
\frac{\sigma}{2}h'(x)^2\right) dx \label{Gten}
\end{equation}
where $h(x)$ is the membrane profile perpendicular to the
adhering cylinder(s), and $\kappa$ is the bending rigidity.
Here, $\frac{1}{2}h'(x)^2$ is the local area increase with
respect to the $x$-$y$ plane. 
The membrane profile $h(x)$ outside of the contact region
with the cylinder(s) has to fulfill the Euler--Lagrange equation
\begin{equation}
 h''''(x) -\xi^2 h''(x)=0 \label{elTen}
\end{equation}
associated with eq.\ (\ref{Gten}). Here,
$\xi=\sqrt{\sigma/\kappa}$ is a characteristic reciprocal length. A
general solution of the Euler--Lagrange equation (\ref{elTen})
is given by
\begin{equation}
h(x) = C_1 + C_2 x + C_3 \exp(-\xi x) + C_4 \exp(\xi x) \label{hTen}
\end{equation}
The tension $\sigma$ is assumed to be constant here. Strictly speaking,
this assumption presupposes a membrane area reservoir, since
the overall area increase of the membrane with respect to the reference plane then depends on 
the distance of the cylinders. 


\subsection{Single cylinder}

\begin{figure}[h]
\begin{center}
\resizebox{\columnwidth}{!}{\includegraphics{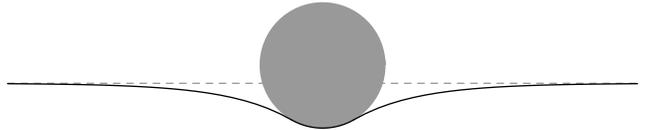}}
\end{center}
\caption{Minimum energy profile for a single cylinder adhering to a 
  membrane under a lateral tension $\sigma$.  The rescaled adhesion
  energy is $\tilde{U}=2 U R^2/\kappa=-2$, and the inverse
  characteristic length $\xi=\sqrt{\sigma/\kappa}$ has the value
  $\xi=1/R$ where $R$ is the cylinder radius.}\label{figProTenOne}
\end{figure}

We first consider a single cylinder adhering to the membrane
in the absence of an external force.
The center of the cylinder is located at $x=0$, and the 
membrane adheres to the cylinder for 
$-\delta_o < x < \delta_o$. The membrane segment in contact
with the cylinder has the circular profile
\begin{equation}
h(x)=h_o - \sqrt{R^2-x^2} + R \simeq h_o + \frac{x^2}{2 R} 
   \label{hTencyl}
\end{equation}
for $\delta_o\ll R$, where $R$ is the cylinder radius.
For $|x|>\delta_o$, the profile of the membrane has the form
\begin{equation} 
h(x)=A +B \exp(-\xi |x|) \label{hTenOut}
\end{equation}
For $A=0$, the profile 
fulfills the boundary condition $h(x)\to 0$ for $|x|\to\infty$. 
The constants $B$ in eq.~(\ref{hTenOut})  and
$h_o$ in eq.~(\ref{hTencyl}) can be determined from the conditions of
continuous profile $h(x)$ and slope $h'(x)$ at $x=\delta_o$. From the latter
condition, we obtain
$B= -\delta_o/(\xi R)$ to first order in $\delta_o$.
The energy $G_z$ of the adhering membrane with profile (\ref{hTencyl}) is
the sum of the elastic energy (\ref{Gten}) and the adhesion energy.
To second order in $\delta_o$, the energy of the adhering membrane
is given by
\begin{equation}
G_z = 2\delta_o\left(\frac{\kappa  }{2R^2} +  U \right)
\label{Gtencyl}
\end{equation}
where $U$ is the adhesion energy per area. The energy of the `outer'
membrane segments with $|x|>\delta_o$ is 
\begin{equation}
G_o=\frac{\sqrt{\sigma\kappa}\delta_o^2}{R^2} \label{GtenOut}
\end{equation}
From the equilibrium condition $\partial G/\partial \delta_o=0$ with
$G=G_z + G_o$,
we obtain
\begin{equation}
\delta_o = -\frac{\kappa + 2 R^2 U}{2\sqrt{\sigma\kappa}} 
     \label{deltaOut}
\end{equation}
The cylinder is bound to the membrane for $\delta_o>0$, i.e.~for
$U<-\kappa/(2 R^2)$. Thus, to obtain a bound state, the
adhesion energy $U$ has to compensate at least
the bending energy $\kappa/ (2 R^2)$ of the membrane with
curvature $1/R$ at the cylinder. Inserting 
(\ref{deltaOut}) into the total membrane energy $G=G_z+G_o$ leads to
\begin{equation}
G = -\frac{(\kappa + 2 R^2 U)^2}{4\sqrt{\sigma\kappa}R^2}
\end{equation}
The profile of a membrane with a single adsorbed cylinder is shown in
Fig.~\ref{figProTenOne}.

\subsection{Two cylinders at same membrane side}

\begin{figure}
\begin{center}
\resizebox{0.8\columnwidth}{!}{\includegraphics{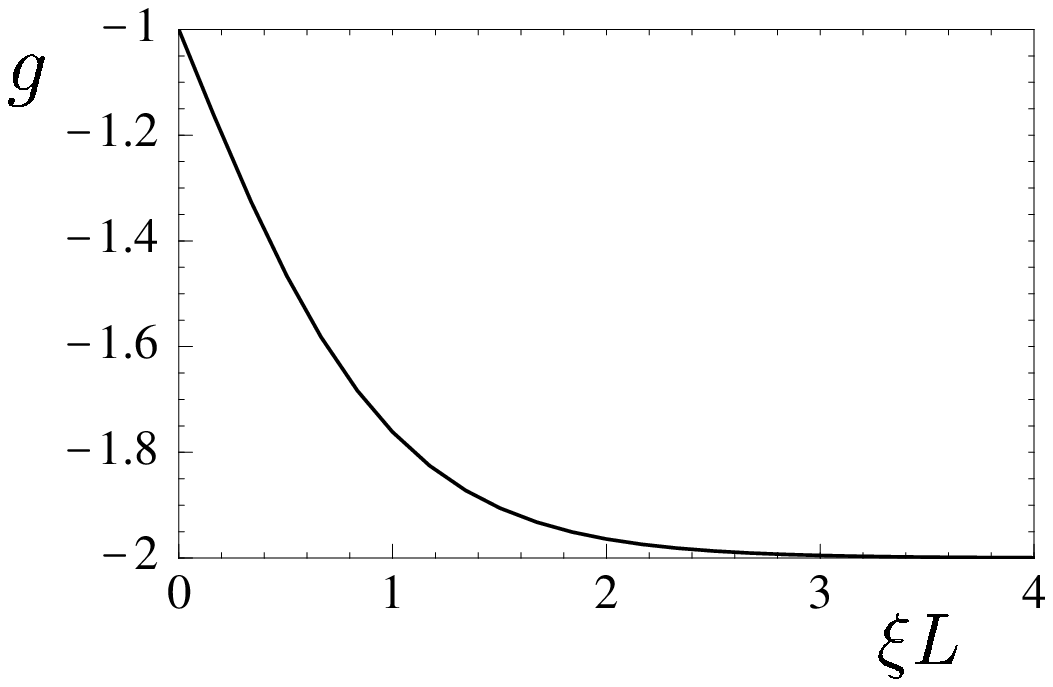}}
\end{center}
\caption{Dimensionless interaction energy $g(L) = -(1+\tanh(\xi L))$
  of two cylinders adhering to the {\em same} side of a membrane under
  lateral tension (see eq.~(\ref{intTenSame})).  Here, $\xi L$ is the
  rescaled distance of the cylinders.
\label{figIntTenSame}}
\end{figure}

We now consider a membrane in contact with two parallel 
cylinders with centers located at $x=\pm L$. In this section, the cylinders
adhere at the same side of the membrane. We assume first
that the membrane is bound to the cylinders for 
$L-\delta_\iota < |x| < L + \delta_o$, and later
determine the contact points $L-\delta_\iota$ and $L + \delta_o$
from a minimization of the energy. A general solution for the shape 
of the `inner' membrane segment  with $|x|< L-\delta_\iota$ obeying the
symmetry condition $h(x) = h(-x)$  is
\begin{equation}
h(x) = C + D\cosh(\xi x) \label{gensolsym}
\end{equation}
From the condition of continuous slope $h'(x)$ at $x=L-\delta_\iota$, we
obtain $D =-\delta_\iota/(\xi R\sinh(\xi L))$
to first order in $\delta_\iota$.
To second order in $\delta_\iota$, the energy of  the `inner' membrane
segment with $-(L-\delta_\iota)< x < L-\delta_\iota$ is 
\begin{equation}
G_\iota = \frac{\sqrt{\sigma\kappa} \delta_\iota^2}{\tanh(\xi L)R^2}
\end{equation}
The total energy of the membrane is $G = G_\iota + G_o + G_z$ with
$G_o$ given in eq.~(\ref{GtenOut}) and $G_z =2
(\delta_\iota+\delta_o)(U + \kappa/(2 R^2))$ as in the previous
section. The contact parameters $\delta_\iota$ and $\delta_o$ follow
from the equilibrium conditions $\partial G/\partial \delta_\iota =0$
and $\partial G/\partial \delta_o=0$. We obtain
\begin{equation}
\delta_\iota=-\frac{(\kappa + 2 R^2 U) \tanh(\xi L)}{2\sqrt{\sigma\kappa}}
\end{equation}
and $\delta_o$ as in eq.~(\ref{deltaOut}), and the interaction energy
\begin{equation}
\fbox{$\displaystyle G(L) = - \frac{(\kappa + 2R^2 U)^2(1+\tanh(\xi L))}
         {4\sqrt{\sigma\kappa} R^2}$} \label{intTenSame}
\end{equation}
The interaction energy is repulsive, attaining its minimum $G
=-(\kappa + 2R^2 U)^2/(2\sqrt{\sigma\kappa} R^2)$ for $L\to \infty$.
The dimensionless interaction energy $g(L) = 
4\sqrt{\sigma\kappa} R^2 G(L)/(\kappa + 2R^2 U)^2$ is shown in
Fig.~\ref{figIntTenSame}.

\subsection{Two cylinders at opposite membrane sides}

\begin{figure}
\begin{center}
\resizebox{0.8\columnwidth}{!}{\includegraphics{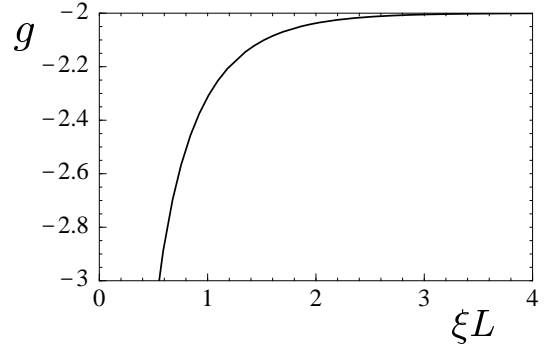}}
\caption{Dimensionless interaction energy $g(L) = -(1+\coth(\xi L)$
  of two cylinders adhering to {\em opposite} sides of a membrane under
  lateral tension (see eq.~(\ref{intTenOpp})).  Here, $\xi L$ is the
  rescaled distance of the cylinders.
\label{figIntTenOpp}}
\end{center}
\end{figure}

In this section, we consider two cylinders adhering at 
opposite sides of the membrane. We assume that 
the `right' cylinder with axis at $x=L$ is on top of the 
membrane, and the `left' cylinder at $x=-L$ below the 
membrane. The membrane segment with
$L-\delta_\iota < x < L+\delta_o$ adhering to the right
cylinder then has the profile
\begin{equation}
h(x) = h_o + \frac{(x-L)^2}{2 R}
\end{equation}
for $\delta_\iota \ll R$ and $\delta_o \ll R$. For symmetry
reasons we now have $h(-x)=-h(x)$. Therefore, the general 
solution for the profile of the `inner' membrane  segment 
with $|x|< L - \delta_\iota$ is
\begin{equation}
 h(x) = C x + D \sinh(\xi x) 
\end{equation}
From the boundary conditions of continuous profile $h$ and
slope $h'$ at $x=L-\delta_\iota$, we obtain 
\begin{eqnarray}
C&=&\frac{\xi R h_o \cosh(\xi L) + \delta_\iota\sinh(\xi L)}
   {R(\xi L  \cosh(\xi L) - \sinh(\xi L))} \\
D&=&\frac{L \delta_\iota + R h_o}
   {R(\sinh(\xi L) -\xi L\cosh(\xi L) )}
\end{eqnarray}
to first order in $h_o$ and $\delta_\iota$. To second order in $h_o$ 
and $\delta_\iota$, the energy of the `inner' membrane segment 
with $|x| < L-\delta_\iota$ is 
\begin{equation}
G_\iota = \frac{\sigma[\xi R^2 h_o^2 \cosh(\xi L)
   + \delta_\iota (L \delta_\iota + 2 R h_o) \sinh(\xi L)]}
{R^2[\xi L \cosh(\xi L) - \sinh(\xi L)]}
\end{equation}
As in the previous sections, the energy $G_o$ of the two `outer' 
segments is given by eq.~(\ref{GtenOut}), and the energy of the
two adhering membrane segments is 
$G_z = 2(\delta_\iota + \delta_o)(U+\kappa/(2 R^2))$.
Minimizing the total membrane energy $G=G_z + G_\iota + G_o$
with respect to $\delta_\iota$, $\delta_o$, and $h_o$ leads to 
\begin{eqnarray}
\delta_\iota &=& -\frac{(\kappa + 2 R^2 U)\coth(\xi L)}
{2\sqrt{\sigma\kappa}} \\
h_o &=& \frac{\kappa + 2 R^2 U}{2\sigma R}
\end{eqnarray}
and $\delta_o$ as in eq.~(\ref{deltaOut}), 
and the attractive interaction energy
\begin{equation}
\fbox{$\displaystyle
G(L) = -\frac{ (\kappa + 2R^2 U)^2(1+\coth(\xi L))}
   {4 \sqrt{\sigma\kappa} R^2}$} \label{intTenOpp}
\end{equation}
The dimensionless energy $g(L) =
4\sqrt{\sigma\kappa} R^2 G(L)/(\kappa + 2R^2 U)^2$ is shown in
Fig.~\ref{figIntTenOpp}.


%
\section{Harmonic potential}

In this section, we assume that the membrane is bound in an external potential, 
induced, e.g., by a substrate supporting the membrane, or an elastic mesh coupled
to the membrane. In harmonic approximation, 
the elastic energy of the membrane then can be written as
\begin{equation}
G =\int \left(\frac{\kappa}{2} h''(x)^2 +
\frac{m}{2}h(x)^2\right) dx \label{Gharm}
\end{equation}
where $m$ is the harmonic potential strength, and $h(x)$ is the membrane 
profile perpendicular to the cylinders.  
We now neglect the lateral tension $\sigma$. This is justified as
long as $\eta\equiv (m/(4\kappa))^{1/4}$,  the characteristic inverse 
length for a membrane in a harmonic potential, is much larger than
$\xi=\sqrt{\sigma/\kappa}$. The Ginzburg-Landau equation reads
\begin{equation}
 h''''(x) +4 \eta^4 h(x)=0 \label{shapeHarmpot}
\end{equation}
%
A general solution of this differential equation is
\begin{eqnarray}
h(x) = C_1 \exp(\eta x) \cos(\eta x)+ C_2 \exp(\eta x) \sin(\eta x) 
     \nonumber \\
       + C_3\exp(-\eta x) \cos(\eta x)+ C_4\exp(-\eta x) \sin(\eta x) 
\end{eqnarray}
\subsection{Single cylinder}

The profile of the membrane segment adhering to the cylinder for
$|x|\le \delta_o$ is again given by eq.~(\ref{hTencyl}). 
To second order in $\delta_o$ and $h_o$, the
energy $G_z$ of this membrane segment is the same as in
eq.~(\ref{Gtencyl}). The harmonic potential contributes a third-order term to 
$G_z$ which is neglected here. 

For $|x|>\delta_o$, the membrane has the form
\begin{equation}
h(x) = A \exp(-\eta |x|) \cos(\eta |x|) +  B \exp(-\eta |x|) \sin(\eta
|x|) \label{hHarmOut}
\end{equation}
which obeys the boundary condition $h(x) \to 0$ for $|x|\to\infty$.
The coefficients $A$ and $B$ follow again from the boundary 
conditions of continuous profile $h(x)$ and slope $h'(x)$ at
$x=\pm \delta_o$. To leading order in $\delta_o$ and $h_o$, these
coefficients are given by $A=h_o + \delta_o/(\eta R)$ and
$B=h_o$.
To second order in $\delta_o$ and $h_o$, the energy of the `outer'
membrane segments with $|x|>\delta_o$ reads
\begin{equation}
G _o=\frac{2\eta\kappa}{R^2} ( \delta_o^2 + 2\eta R \delta_o
h_o + 2
\eta^2 R^2 h_o^2)
 \label{GharmOut}
\end{equation}
The contact point $\delta_o$ and height $h_o$ follow from the equilibrium
conditions $\partial G/\partial h_o = 0$ and 
$\partial G/\partial \delta_o = 0$ where $G=G_z + G_o$ is the total
membrane energy. We obtain
\begin{eqnarray}
\delta_o  &=& - \frac{\kappa + 2 R^2 U}{2\eta\kappa} \label{deltaOutHarm}\\
h_o &=&  \frac{\kappa + 2 R^2 U}{4\eta^2\kappa R}  \label{hoEins}
\end{eqnarray}
and  the energy
\begin{equation}
G= - \frac{(\kappa + 2 R^2 U)^2}{4\eta\kappa R^2}
\end{equation}

If an external force $F$ is acting on the cylinder perpendicular
to the membrane plane, the total energy is $G=G_z + G_o + F
h_o$. Minimizing $G$ now leads to 
\begin{eqnarray}
\delta_o = \frac{F R - 2\eta(\kappa + 2 R^2 U)}{4\eta^2\kappa}\\
h_o=\frac{-F R + \eta(\kappa + 2 R^2 U)}{4\eta^3\kappa R}
\end{eqnarray}
The cylinder unbinds from the membrane if the contact parameter 
$\delta_o$, and thus the contact area, equals zero.
The threshold force leading  to unbinding  is
\begin{equation}
F_t = 2\eta(\kappa + 2 R^2 U)/R
\end{equation}

\subsection{Two cylinders at same membrane side}

\begin{figure}
\begin{center}
\resizebox{\columnwidth}{!}{\includegraphics{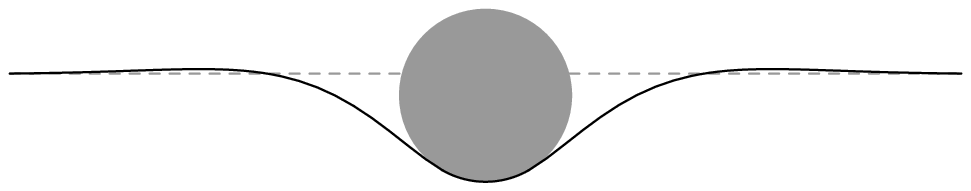}}

\vspace{1.2cm}

\resizebox{\columnwidth}{!}{\includegraphics{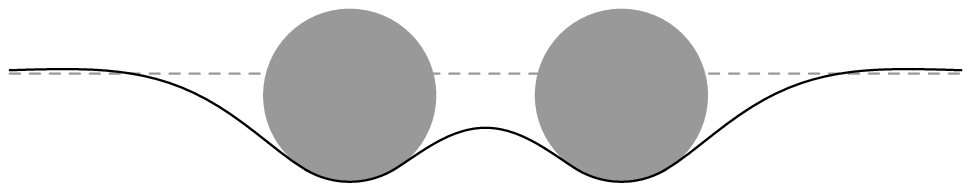}}
\end{center}
\caption{Minimum energy profiles for (top) a single adsorbed 
cylinder and (bottom) two cylinders adhering at the same side
of a membrane in a harmonic potential with strength $m$. The rescaled 
adhesion energy is $\tilde{U}=2 U R^2/\kappa=-2$, and the
inverse characteristic length  $\eta=(m/(4\kappa))^{1/4}$ has
the value $\eta=1/R$ where $R$ is the cylinder radius.
}
\label{profilesHarm}     
\end{figure}

\begin{figure}
\begin{center}
\resizebox{0.8\columnwidth}{!}{\includegraphics{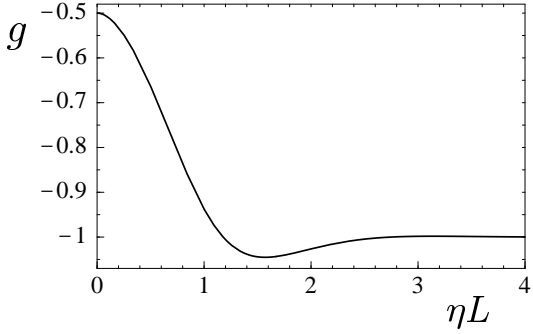}}
\caption{Dimensionless interaction energy 
  $g(L) = -\exp(2\eta L)/(\exp(2\eta L)+\cos(2\eta L)+\sin(2\eta L))$
  of two cylinders adhering to the {\em same} side of a membrane in
  a harmonic potential (see eq.~(\ref{intHarmSame})).  
\label{figIntHarmSame}}
\end{center}
\end{figure}

Let us now consider a membrane with two cylinders adhering at the
same side of the membrane. For 
symmetry reasons, the shape of the `inner' membrane segment
with $|x|< L-\delta_\iota$ is given by
\begin{equation}
h(x) = C \sinh(\eta x) \sin(h x) + D \cosh(\eta x)\cos(\eta x)
\end{equation}
The coefficients are again obtained from the boundary 
conditions of continuous profile $h(x)$ and slope $h'(x)$ at
$x=\pm (L-\delta_\iota)$. To first order in $\delta_\iota$ and
$h_o$ they read
\begin{eqnarray}
C &=& \frac{2}{\eta R \Phi_1} \big\{[ \eta R h_o \sin(\eta L)
-\delta_\iota \cos(\eta L)  ]\cosh(\eta L) \nonumber \\
&&\hspace*{1.5cm}-\eta R h_o\cos(\eta L)\sinh(\eta L)\big\} 
\\
D &=& \frac{2}{\eta R \Phi_1} \big\{ [\eta R h_o \cos(\eta L)
 +\delta_\iota \sin(\eta L)]\sinh(\eta L) \nonumber\\
&&\hspace*{1.5cm}+\eta R h_o \cosh(\eta L)\sin(\eta L)\big\}
\end{eqnarray}
with
$\Phi_1 = \sin(2 \eta L) + \sinh(2\eta L)$.
To second order in $\delta_\iota$ and $h_o$, the energy of the inner 
membrane segment is given by
\begin{eqnarray}
G_{\iota} &=& \frac{2\eta\kappa}{R^2 \Phi_1} \bigg\{ 
  (\delta_\iota^2 - 2 \eta^2 R^2 h_o^2) \cos(2\eta L) 
   + (\delta_\iota^2 + 2 \eta^2 R^2 h_o^2)\times  \nonumber \\
  && \hspace*{-0.3cm}\times \cosh(2\eta L)
  + 2\eta R \delta_\iota h_o[\sinh(2\eta L) - \sin(2\eta L) ]\bigg\}
\end{eqnarray}
Minimizing the total membrane energy $G= G_\iota + G_o + G_z$ 
with $G_z = 2(\delta_\iota + \delta_o)( U+\kappa/(2R^2))$ 
and $G_o$ as given in eq.~(\ref{GharmOut}) with respect to
$\delta_\iota$, $\delta_o$, and
$h_o$ leads to 
\begin{eqnarray}
\delta_\iota &=& -\frac{(\kappa + 2 R^2 U)(\exp(2 \eta L)-\cos(2\eta
L))}
   {2\eta\kappa(\exp(2\eta L)+\cos(2\eta L) + \sin(2\eta L))}\\
\delta_o &=& -\frac{(\kappa + 2 R^2 U)(\exp(2\eta L)+\cos(2\eta L))}
   {2\eta\kappa(\exp(2\eta L)+\cos(2\eta L) + \sin(2\eta L))}\\
h_o &=& \frac{(\kappa + 2 R^2 U)(\exp(2\eta L)+\cos(2\eta
L)-\sin(2\eta L))}
   {4\eta^2\kappa R(\exp(2\eta L)+\cos(2\eta L) + \sin(2\eta
L))}\hspace{0.8cm}
\end{eqnarray}
and the interaction energy
\begin{equation}
\fbox{$\displaystyle
G(L) = -\frac{(\kappa + 2 R^2 U)^2 \exp(2\eta L)}
{2\eta\kappa R^2[ \exp(2\eta L) + \cos(2\eta L) + \sin(2\eta L)]}$} \label{intHarmSame}
\end{equation}
The dimensionless energy 
$g(L)= 2\eta\kappa R^2 G(L) / (\kappa + 2 R^2 U)^2$ is shown 
in Fig.~\ref{figIntHarmSame}. The global minimum of the interaction 
energy is at $L_{\text{min}}=\pi/(2 \eta)$ (see also Fig.~\ref{profilesHarm}).

\subsection{Two cylinders at opposite membrane sides}

\begin{figure}[t]
\begin{center}
\resizebox{0.8\columnwidth}{!}{\includegraphics{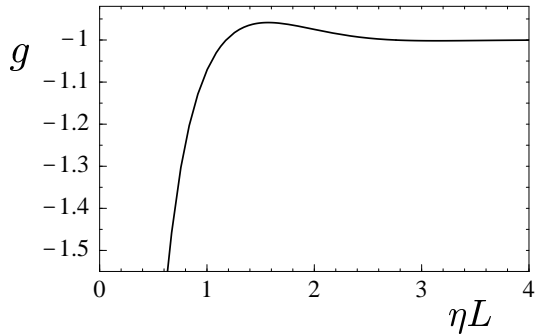}}
\caption{Dimensionless interaction energy 
  $g(L) = -\exp(2\eta L)/(\exp(2\eta L)-\cos(2\eta L)-\sin(2\eta L))$
  of two cylinders adhering to {\em opposite} sides of a membrane in
  a harmonic potential (see eq.~(\ref{intHarmOpp})).  
\label{figIntHarmOpp}}
\end{center}
\end{figure}

In this section, we consider two cylinders adhering at opposite
sides of a membrane in a harmonic potential.
As in section 3.3, we assume that the `right' cylinder with axis
at $x=L$ is above the membrane, and the left cylinder at 
$x=-L$ is below the membrane.
The profile of the membrane has the symmetry
$h(-x)=-h(x)$. Therefore, the profile of the 
  `inner' membrane segment
with $|x|< L-\delta_\iota$ has the general form
\begin{equation}
h(x) = C \cosh(\eta x) \sin[h x] + D \sinh(\eta x)\cos(\eta x)
\end{equation}
From the 
conditions of continuous profile $h(x)$ and slope $h'(x)$ at
$x=\pm (L-\delta_\iota)$, we obtain to first order in 
$\delta_\iota$ and $h_o$:
\begin{eqnarray}
C &=& \frac{2}{\eta R \Phi_2} \big\{ \eta R h_o[
(\cosh(\eta L)\cos(\eta L) -\sinh(\eta L)\sin(\eta L)]
 \nonumber \\
&&\hspace*{1.5cm}+\delta_\iota\sinh(\eta L)\cos(\eta L)\big\} 
\\
D &=& -\frac{2}{\eta R \Phi_2} \big\{ \eta R h_o[
(\cosh(\eta L)\cos(\eta L) +\sinh(\eta L)\sin(\eta L)]
 \nonumber \\
&&\hspace*{1.5cm}+\delta_\iota \cosh(\eta L)\sin(\eta L)\big\}
\end{eqnarray}
with $\Phi_2 = \sin(2 \eta L) -\sinh(2\eta L)$.
To second order in $\delta_\iota$ and $h_o$, the energy of the inner 
membrane segment is given by
\begin{eqnarray}
G_{\iota} &=& \frac{2\eta\kappa}{R^2 \Phi_2} \bigg\{ 
  (\delta_\iota^2 - 2 \eta^2 R^2 h_o^2) \cos(2\eta L) 
   - (\delta_\iota^2 + 2 \eta^2 R^2 h_o^2)\times  \nonumber \\
  && \hspace*{-0.3cm}\times \cosh(2\eta L)
  - 2\eta R \delta_\iota h_o[\sinh(2\eta L) + \sin(2\eta L) ]\bigg\}
\end{eqnarray}
Minimizing the total membrane energy $G= G_\iota + G_o + G_z$, 
with 
$G_z = 2(\delta_\iota + \delta_o)( U+\kappa/(2R^2))$ 
and $G_o$ given in eq.~(\ref{GharmOut})  leads to 
\begin{eqnarray}
\delta_\iota &=& -\frac{(\kappa + 2 R^2 U)(\exp(2 \eta L)+\cos(2\eta
L))}
   {2\eta\kappa(\exp(2\eta L)-\cos(2\eta L) - \sin(2\eta L))}\\
\delta_o &=& -\frac{(\kappa + 2 R^2 U)(\exp(2\eta L)-\cos(2\eta L))}
   {2\eta\kappa(\exp(2\eta L)-\cos(2\eta L) - \sin(2\eta L))}\\
h_o &=& \frac{(\kappa + 2 R^2 U)(\exp(2\eta L)-\cos(2\eta
L)+\sin(2\eta L))}
   {4\eta^2\kappa R(\exp(2\eta L)-\cos(2\eta L) - \sin(2\eta
L))}\hspace{0.8cm}
\end{eqnarray}
and the interaction energy
\begin{equation}
\fbox{$\displaystyle
G(L) = -\frac{(\kappa + 2 R^2 U)^2 \exp(2\eta L)}
{2\eta\kappa R^2[ \exp(2\eta L) - \cos(2\eta L) - \sin(2\eta L)]}$}
\label{intHarmOpp}
\end{equation}
The dimensionless interaction energy $g(L)= 2\eta\kappa R^2 G(L)/
(\kappa + 2 R^2 U)^2$ is shown in Fig.~\ref{figIntHarmOpp}.

\subsection{External force}

In the presence of an external force $F$ acting on the cylinders perpendicular
to membrane plane, the
total energy for given contact parameters $\delta_\iota$ and
$\delta_o$ and given $h_o$ is $G=G_z(\delta_\iota,\delta_o) +
G_\iota(\delta_\iota,h_o) + G_o(\delta_o,h_o) + 2 F h_o$. Minimizing
$G$ with respect to $\delta_o$, $\delta_\iota$, and $h_o$ leads to
\begin{eqnarray}
G(L)&=&-\frac{(\kappa+2R^2 U)^2}{2\eta\kappa R^2 \Phi_3}
\bigg\{e^{2\eta L} + \tilde{F}\big[e^{2\eta L} +\cos(2\eta L)
\nonumber\\  
&& \hspace*{0.1cm} -\sin(2\eta L)\big] 
+\tilde{F}^2[\cosh(2\eta L)+\cos(2\eta
L)]\bigg\}\hspace{0.5cm}
\end{eqnarray}
with $\Phi_3 = \exp(2\eta L) +\cos(2\eta L) +\sin(2\eta L)$ and
the rescaled force $\tilde{F}=-F R/[\eta(\kappa+2 R^2 U)]$.

\begin{figure}
\begin{center}
\resizebox{0.8\columnwidth}{!}{\includegraphics{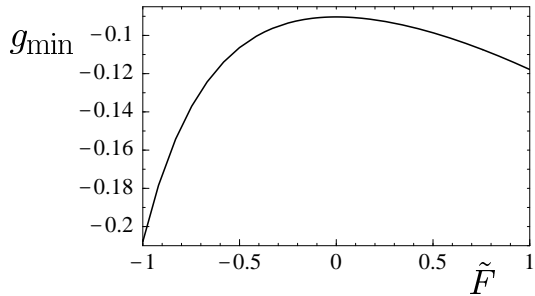}}
\caption{Dimensionless binding energy $g_{\text{min}}$ as a function
of the rescaled force $\tilde{F}$ for  two cylinders
adhering at the same side of a membrane in a harmonic potential.
The absolute value of the binding energy  $g_{\text{min}}$ 
{\it increases} under an applied force $\tilde{F}$, irrespective of the
sign of $\tilde{F}$.
\label{figIntHarmForce}}
\end{center}
\end{figure}

Fig.~\ref{figIntHarmForce} shows the dimensionless binding energy
of the cylinders $g_{\text{min}} = 
(G(L_{\text{min}}) - G(\infty))\times (2\eta\kappa R^2)/(\kappa+2R^2 U)^2$ as a function of the 
rescaled force $\tilde{F}$. Here, $L_{\text{min}}$ is the cylinder separation at which the
interaction energy $G(L)$ has its minimum. The binding energy of the two cylinders
increases irrespective of the sign of the force, i.e.~both for perpendicular forces 
pulling on the cylinders, and for forces pushing the cylinders  into the membrane.

\section{Discussion}

We have considered the indirect interactions of parallel cylindrical 
objects adhering to a planar membrane. The interactions arise from
the perturbations of the equilibrium membrane shape caused by
the cylinders. The cylinders are assumed to be much longer than their 
distance $L$, and the membrane profile  is approximated
by a one-\-dimen\-sional function $h(x)$. Here, $x$ is a cartesian 
coordinate perpendicular to the cylinder axes. 

For a pair of cylinders adhering to a membrane under lateral tension, the 
indirect interaction is repulsive if the cylinders adhere to the same membrane 
side, and attractive for two cylinders adhering at opposite membrane
sides (see section 3). 
For a membrane in a harmonic potential, the interaction of the cylinders
is always attractive (see section 4). The latter situation applies, in first
approximation, to supported or bound membranes. 

 These results are intuitively plausible:
The curvature $h''(x)$ around a single cylinder adhering to the top of
a membrane under lateral tension is negative (see eq.~(\ref{hTenOut}) and 
Fig.~\ref{figProTenOne}). This means that the 
membrane is `curved away' from a second cylinder approaching the membrane 
from the same side, but `curved towards' a cylinder approaching the
membrane from the opposite side. The negative curvature makes it more
difficult for a cylinder to adhere at the same side close to the first cylinder,
but facilitates the adhesion of a second cylinder at the opposite side.
In contrast, the profile around a single cylinder adhering to a membrane
in a harmonic potential is an exponentially damped, oscillating profile,
with both negatively and positively curved segments 
(see eq.~(\ref{hHarmOut}) and Fig.~\ref{profilesHarm}).

For a membrane under lateral tension, the profile around a
single adhering cylinder with axis at $x=0$ is proportional 
to $\exp(-\xi |x|)$ (see section 3.1). Here, $\xi =\sqrt{\sigma/\kappa}$ 
is the characteristic reciprocal length of a membrane with
tension $\sigma$. The same profile is obtained around
a single wedge-shaped inclusion in a tense membrane 
(see appendix B.1). Similar membrane profiles are also obtained
for a membrane with two cylinders or two wedge-shaped inclusions.
However, the contact region of the
adsorbed cylinders depends on the distance $L$ of the
cylinders, whereas the angle
$\alpha$ and width $D$ of the wedge-shaped inclusions are constant.
Therefore, the interaction energy as a function of the distance
$L$ is different for adsorbed cylinders
and for wedge-shaped inclusions 
(see eqs.~(\ref{intTenSame}), (\ref{intTenOpp}), (\ref{wedgesOne}), 
and (\ref{wedgesTwo})). For the same reason the
interaction energy of adhering spheres should be different
from the interaction energy of conical inclusions.

The strength of the cylinder interactions can be estimated from the energetic prefactors.
For a membrane under lateral tension, the prefactor 
$(\kappa + 2 R^2 U)^2/(4\sqrt{\sigma\kappa}R^2)$ of the interaction energies (\ref{intTenSame})
and (\ref{intTenOpp}) can be written as $(\delta_o/R)^2 \sqrt{\sigma\kappa}$, where 
$\delta_o$ is the contact point at large separations $L$ (see eq.~\ref{deltaOut}).
For $(\delta_o/R)\lesssim 0.5$, the typical bending rigidity $\kappa=10^{-19}$ J and membrane tensions up to 
$10^{-3}$ N/m, the prefactor can attain values of up to 0.5 $k_B T$ per nanometer length of the cylinders, leading to large interaction energies for colloidal cylinders with lengths of
several hundred nanometers. 
For a membrane in an external  potential, the prefactor $(\kappa + 2 R^2 U)^2/(2\eta\kappa R^2)$
of the interaction energies (\ref{intHarmSame}) and (\ref{intHarmOpp}) can be written in the form $2(\delta_o/R)^2 \eta\kappa$ with $\delta_o$ given by eq.~(\ref{deltaOutHarm}). For
$\delta_o/R=0.5$,  a typical length scale $1/\eta = 50$ nm of about 10 times the membrane thickness, and $\kappa=10^{-19}$ J, this prefactor has the value
0.25 $k_B T$/nm. The binding energy for two cylinders adhering at the same side then is
about -0.025 $k_B T$ per nanometer length of the cylinders (see Fig.~\ref{figIntHarmSame}), leading, e.g., to a binding energy of 10 $k_B T$ for colloidal cylinders with a length of 400 nm. 
These estimates for the static interactions of the 
colloidal membrane-adsorbed cylinders are clearly
larger than the loss of orientational free energy of order $k_B T$ per cylinder, which is caused by the
parallel alignment. The interactions therefore
can lead to stable pairs or bundles of such cylinders. However, for a membrane dominated by lateral tension, these conformations are only stable if the cylinders are adsorbed alternately on opposite membrane sides.

Attractive interactions and aggregation of spherical particles 
adsorbed on vesicles has been experimentally observed by Koltover, R\"adler, and
Safinya \cite{koltover99}. The particles do not aggregate in solution. Therefore,
the attractive interactions are probably indirect, i.e.~mediated by the 
membrane. Since the particles have a radius of 0.9 $\mu m$, the interactions are
most likely due to the rather long-ranged fluctuation-induced interactions,
or interactions from perturbations of the equilibrium membrane shape as
considered here. Interactions due to membrane thickness perturbations
typically decay over a few nanometers, a length-scale comparable to the
membrane thickness \cite{dan94,fournier98,may99,harroun99,sens00,schiller00} . 
The fluc\-tua\-tion-induced pair 
interactions are always attractive, but rather weak compared to thermal
energies \cite{goulian93,netz,park96,golestanian96,dommersnes99,helfrich01}.
 Depending on the induced deformation, the static interactions
considered in this paper can be strong, see above. Exploring the latter as a possible
explanation for the observations made by Koltover et al.\
requires an extension of the calculations presented here to beads
and to nonplanar membranes or vesicles. 

\vspace{1cm}

\noindent
{\bf \large Acknowledgements}\\[0.2cm]
I would like to thank Wolfgang Helfrich and Martin Brink\-mann for helpful discussions.

\begin{appendix}

\section{Single cylinder adhering to a finite membrane
under lateral tension}

In this appendix, we consider a single cylinder adhering to a {\em finite}
membrane under lateral tension. In the limit of infinite membrane
size, we will obtain and thus corroborate the profile and energy
derived in section 3.1. 

As in section 3.1, the membrane segment 
in contact  with the cylinder for $-\delta_o<x<\delta_o$ has the 
circular profile (\ref{hTencyl}). The profile $h(x)$ of the nonadhering
finite membrane segment for $\delta_o<x<\Lambda$ has the general form
(\ref{hTen}) for $x>0$. For symmetry reasons, we have $h(-x)=h(x)$. The four 
parameters $C_1$ to $C_4$ in (\ref{hTen})
can be determined from the four boundary conditions:
\begin{eqnarray}
h(\delta_o) = h_o + \frac{\delta_o^2}{2 R}\; , \;\;\;  h(\Lambda) = 0 \\
h'(\delta_o) = \frac{\delta_o}{R}\; , \;\;\;  h'(\Lambda) = 0 \\
\end{eqnarray}
Here, we again assume $\delta_o\ll R$, and determine $C_1$ to 
$C_4$ to first order in $\delta_o$ and $h_o$. The second order
in $\delta_o$ and $h_o$, the elastic energy of the two outer
membrane segments is
\begin{eqnarray}
G_o= \sigma\bigg[ C_2^2 L +
C_3^2\xi (1-e^{-2\xi \Lambda}) +
C_4^2\xi (e^{2\xi \Lambda}-1) \nonumber \\
+ 2 C_2 C_3 (e^{-\xi \Lambda}-1) + 2 C_2 C_4 (e^{\xi \Lambda}-1)\bigg]
\end{eqnarray}
For given $h_o$ and $\delta_o$, the analytical expressions for
the four parameters $C_1$ to $C_4$ are rather lengthy and not
shown here. However, if we minimize the total free energy 
$G=G_z + G_o$ with $G_z$ as in eq.~(\ref{Gtencyl}) with
respect to $h_o$ and $\delta_o$, we get
\begin{eqnarray}
C_1 &=& -\frac{(\kappa + 2 R^2 U)}{2\sigma R\cosh(\xi\Lambda)}\\
C_2 &=&0\\
C_3 &=& \frac{(\kappa + 2 R^2 U)(\tanh(\xi\Lambda) + 1)}{4\sigma R}\\
C_4 &=& -\frac{(\kappa + 2 R^2 U)(\tanh(\xi\Lambda) - 1)}{4\sigma R}
\end{eqnarray}
For $\Lambda\to\infty$, we obtain $C_1\to 0$, 
$C_3 \to(\kappa + 2 R^2 U)/(2\sigma R)$, and 
$C_4 \exp(\xi x)\to 0$. In the limit of infinite membrane size, we thus
recover the equilibrium shape 
$h(x)= (\kappa + 2 R^2 U)\exp(-\xi x)/ (2\sigma R)$ from section 3.1.

\section{Wedge-shaped inclusions}
\subsection{Single inclusion}

A wedge-shaped inclusion is characterized by its diameter $D$ and angle $\alpha$.
If the center of the inclusion is located at $x=0$, a single inclusion imposes
the boundary condition 
\begin{equation}
h'(\pm D/2) = \pm \alpha \label{cone}
\end{equation} 
For a membrane under lateral tension, the profile around a single inclusion
is given by
\begin{equation}
h(x) = A + B\exp(-\xi |x|) \label{proConeOne}
\end{equation}
as in the case of a single adhering cylinder (see section 2.1).  From the
boundary  condition (\ref{cone}), one obtains
\begin{equation}
B = -\alpha \exp( \xi D/2)/\xi \label{hconout}
\end{equation}
The coefficient $A$ in (\ref{proConeOne}) is arbitrary and defines the
reference plane.
The elastic energy of the membrane around the inclusion is given by
\begin{equation}
G = \alpha^2 \sqrt{\sigma\kappa}
\end{equation}

\subsection{Two inclusions with equal orientation} 

We now consider two inclusions of diameter $D$ with `inner' edges 
located at $x=\pm L$, and `outer' edges located at $x=\pm (L+D)$.
In this section, the two inclusions have the same orientation.
At their edges, the inclusions impose the boundary conditions
\begin{eqnarray}
h'(\pm L) = \pm (\beta -\alpha) \label{innerEdge}\\
h'(\pm (L+D)) = \pm (\beta + \alpha) \label{outerEdge}
\end{eqnarray}
on the membrane where $\beta$ is the tilt angle of the inclusions. The
membrane profile is symmetric with $h(-x) = h(x)$, and decays to zero for
$x\to\infty$. The profile of the `inner' membrane segment between the
inclusions is given by
\begin{equation}
h(x) = C + D \cosh(\xi x)
\end{equation}
From the boundary condition (\ref{innerEdge}), one obtains 
\begin{equation}
D = \frac{\beta-\alpha}{\xi \sinh(\xi L)}
\end{equation}
The energy of the `inner' membrane segment then is given by
\begin{equation}
G_\iota = (\beta - \alpha)^2 \sqrt{\sigma \kappa} \coth(\xi L)
\end{equation}

The `outer' membrane segments for $|x| > (L + D)$ 
have again the shape (\ref{hconout}). The boundary condition
(\ref{outerEdge}) is fulfilled for
\begin{equation}
B = -(\beta+\alpha) \exp( \xi (L+D))/\xi 
\end{equation}
and the elastic energy of the two `outerÔ segments is 
\begin{equation}
G_o = (\alpha + \beta)^2 \sqrt{\sigma\kappa} \label{GoCones}
\end{equation}
As in the previous section, the coefficient $A$ in (\ref{proConeOne}) 
is arbitrary.
Minimizing the total membrane energy $G = G_\iota + G _o$ with
respect to the tilt angle $\beta$ of the inclusions leads to
the equilibrium value
\begin{equation}
\beta = \alpha \exp(-2\xi L)
\end{equation}
and the interaction energy
\begin{equation}
\fbox{$\displaystyle
G(L) = 2 \alpha^2 \sqrt{\sigma\kappa} (1 + \exp(-2\xi L)) $}
\label{wedgesOne}
\end{equation}
\subsection{Two inclusions with opposite orientation} 

If the two inclusions have opposite orientation with respect to the
membrane plane, the boundary conditions at the inclusion edges
are
\begin{eqnarray}
h'(\pm L) =  (\beta -\alpha) \label{innerEdge2}\\
h'(\pm (L+D)) =  (\beta + \alpha) \label{outerEdge2}
\end{eqnarray}
and
\begin{eqnarray}
h(\pm L) = \pm h_o 
\end{eqnarray}
where $h_o$ is the deviation of the inclusions out of the reference plane,
and $\beta$ is again the tilt angle. We now have the symmetry 
$h(-x) =-h(x)$.  The profile of the `inner' 
membrane segment between the inclusions is given by
\begin{equation}
h(x) = C x+ D \sinh(\xi x)
\end{equation}
From the boundary conditions above, we obtain 
\begin{eqnarray}
C &=& \frac{\xi h_o\cosh(\xi L) + (\alpha-\beta)\sinh(\xi L)}
        {\xi L\cosh(\xi L) - \sinh(\xi L)} \\
D &=& -\frac{h_o + (\alpha - \beta) L}
  {\xi L\cosh(\xi L) - \sinh(\xi L)} 
\end{eqnarray}
The energy of the `inner' membrane segment  is 
\begin{equation}
G_\iota = \frac{\sigma}{2}\left( 2 C^2 L + 4 C D \sinh(\xi L)
 + D^2 \xi \sinh(2\xi L)\right)
\end{equation}
with $C$ and $D$ as given above. As in the previous section, the
two outer membrane segments have the energy (\ref{GoCones}).

Minimizing the total membrane energy $G = G_\iota + G _o$ with
respect to the tilt angle $\beta$ and deviation $h_o$ of the inclusions
leads to the equilibrium values
\begin{eqnarray}
\beta &=& -\alpha \exp(-2\xi L) \\
h_o &=& -\alpha (1-\exp(-2\xi L))/\xi
\end{eqnarray}
and the attractive interaction energy
\begin{equation}
\fbox{$\displaystyle
G(L) = 2 \alpha^2 \sqrt{\sigma\kappa} (1 - \exp(-2\xi L)) $}
\label{wedgesTwo}
\end{equation}

\end{appendix}


\begin{thebibliography}{99}

\bibitem{alberts94} B.~Alberts et al., {\it Molecular Biology of the Cell},
3rd ed.\ (Garland, New York, 1994).

\bibitem{lipowsky95} R.~Lipowsky and E.~Sackmann, 
 {\it The Structure and Dynamics of Membranes} 
(Elsevier, Amsterdam, 1995).

\bibitem{simons97} K.~Simons and E.~Ikonen, Nature {\bf 387}, 
    569 (1997).

\bibitem{schekman96} R.~Schekman and L.~Orci, Science {\bf 271}, 1526
(1996).

\bibitem{monks98} C.R.F.~Monks, B.A.~Freiberg, H.~Kupfer, N.~Sciaky, 
   and A.~Kupfer, Nature {\bf 395}, 82 (1998). 

\bibitem{grakoui99} A.~Grakoui, S.K.~Bromley, C.~Sumen, M.M.~Davis,
A.S.~Shaw, P.M.~Allen, and M.L.~Dustin, Science {\bf 285}, 221 (1999).

\bibitem{keller98} S.L.~Keller, W.H.~Pitcher III, W.H.~Huestis, 
H.M.~McConnell, Phys.~Rev.~Lett.\ {\bf 81}, 5019 (1998).

\bibitem{dietrich01} C.~Dietrich, L.A.~Bagatolli, Z.N.~Volovyk, 
N.L.~Thompson, M.~Leve, K.~Jacobson, and E.~Gratton,
Biophys.~J.\ {\bf 80}, 1417 (2001).

\bibitem{sackmannGroup}
A.~Albersd\"orfer, T.~Feder, and E.~Sackmann, Biophys. J. {\bf 73},  245
  (1997); A.~Kloboucek, A.~Behrisch, J.~Faix, and E.~Sackmann,
    Biophys.~J.~{\bf 77}, 2311 (1999); 
Z.\ Guttenberg, B.\ Lorz, E.\ Sackmann, and
A.\ Boulbitch, Europhys.\ Lett.\ {bf 54}, 826 (2001).

\bibitem{koltover99} I.\ Koltover, J.O.\ R\"adler, and 
    C.R.\ Safinya, Phys.\ Rev.\ Lett.\ {\bf 82}, 1991 (1999).

\bibitem{goulian93} M.\ Goulian, R.\ Bruinsma, and P.\ Pincus,
   Europhys.\ Lett.\ {\bf 22}, 145 (1993); 
    Erratum in  Europhys.\ Lett.\ {\bf 23}, 155 (1993).

\bibitem{netz} R.R.\ Netz and P.\ Pincus, 
Phys.\ Rev.\ E {\bf 52}, 4114 (1995);
R.R.~Netz, J.\ Phys.\ I France {\bf 7}, 833 (1997).

\bibitem{park96} J.-M.\ Park and T.C.\ Lubensky, 
J.\ Phys.\ I France {\bf 6}, 1217 (1996).

\bibitem{golestanian96} R.\ Golestanian, M.\ Goulian, and M.\ Kardar,
     Europhys.~Lett.~{\bf 33}, 241 (1996);
  Phys.\ Rev.\ E.\ {\bf 54}, 6725 (1996).

\bibitem{dommersnes99} P.G.\ Dommersnes and J.-B.\ Fournier,
Europhys.\ Lett.\ {\bf 46}, 256 (1999);
Eur.\ Phys.\ J.\ B {\bf 12}, 9 (1999)

\bibitem{helfrich01} W.\ Helfrich and T.R.\ Weikl,
Eur.\ Phys.\ J.\ E {\bf 5}, 423 (2001).

\bibitem{tw1} T.R.\ Weikl, Europhys.\ Lett.\ {\bf 54}, 547 (2001);
 Phys.\ Rev.\ E.\ 66, 061915 (2002).

\bibitem{bruinsma94} R.\ Bruinsma, M.\ Goulian, and P.\ Pincus,
    Biophys.\ J.\ {\bf 67}, 746 (1994).

\bibitem{tw2} T.R.\ Weikl, R.R.\ Netz, and R.\ Lipowsky, 
    Phys.\ Rev.\ E {\bf 62}, R45 (2000);
   T.R.\ Weikl and R.\ Lipowsky, Phys.\ Rev.\ E {\bf 64}, 011903 (2001).

\bibitem{dan94} N.\ Dan, A.\ Berman, P.\ Pincus, and S.\ Safran,
     J.\ Phys.\ II France {\bf 4}, 1713 (1994).

\bibitem{fournier98} J.-B.\ Fournier, 
Europhys.\ Lett.\ {\bf 43}, 725 (1998). 

\bibitem{may99} S.\ May and A.\ Ben-Shaul,
Biophys.\ J.\ 76, 751 (1999).

\bibitem{harroun99} T.A.~Harroun, W.T.~Heller, T.M.~Weiss, L.~Yang,
and H.W.~Huang, Biophys.~J.\ {\bf 76}, 937 (1999).

\bibitem{sens00} P.\ Sens and S.A.\ Safran, Eur.\ Phys.\ J.\ E {\bf 1}, 237 (2000).

\bibitem{schiller00} P.~Schiller, Phys.\ Rev.\ E {\bf 62}, 918 (2000);
Mol.\ Phys.\ {\bf 98}, 493 (2000).

\bibitem{tw98} T.R.\ Weikl, M.M.\ Kozlov, and W.\ Helfrich,
Phys.\ Rev.\ E.\ {\bf 57}, 6988 (1998).

\bibitem{dommersnes98} P.G.\ Dommersnes, J.-B.\ Fournier, and
P.\ Galatola, Europhys.\ Lett.\ {\bf 42}, 233 (1998).

\bibitem{kim98_99} K.S.\ Kim, J.\ Neu, and G.\ Oster,
Biophys.\ J.\ {\bf 75}, 2274 (1998); 
Europhys.\ Lett.\ {\bf 48}, 99 (1999).

\bibitem{sintes98} T.\ Sintes, A.\ Baumg\"artner,
J.\ Phys.\ Chem.\ B {\bf 102}, 7050 (1998).

\bibitem{dommersnes02} P.G.\ Dommersnes and J.-B.\ Fournier,
Biophys.\ J.\ {\bf 83}, 2898 (2002). 

\bibitem{breidenich00} M.\ Breidenich, R.R.\ Netz, and R.\ Lipowsky,
Europhys.\ Lett.\ {\bf 49}, 431 (2000).

\bibitem{bickel00_01} T.\ Bickel, C.\ Marques, and C.\ Jeppesen,
Phys.\ Rev.\ E {\bf 62}, 1124 (2000); 
T.\ Bickel, C.\ Jeppesen, and C.M.\ Marques,
      Eur.\ Phys.\ J.\ E {\bf 4}, 33 (2001).

\bibitem{boulbitch02} A.\ Boulbitch, Europhys.\ Lett.\ {\bf 59}, 910-915 (2002).

\end{thebibliography}
\end{document}